\documentclass{article}

\PassOptionsToPackage{numbers, compress}{natbib}

\usepackage[final]{neurips_2022_ml4ps}
\newcommand{\fink}{\texttt{Fink}}
\newcommand{\elasticc}{\texttt{ELAsTiCC}}
\newcommand{\elasticcd}{\texttt{ELAsTiCC\_data}}

\newcommand{\simbad}{\texttt{SIMBAD}}
\newcommand{\tns}{\texttt{TNS}}
\newcommand{\finkd}{\texttt{Fink-data}}

\usepackage[utf8]{inputenc} 
\usepackage[T1]{fontenc}    
\usepackage{hyperref}       
\usepackage{url}            
\usepackage{booktabs}       
\usepackage{amsfonts}       
\usepackage{nicefrac}       
\usepackage{microtype}      
\usepackage{xcolor}         
\usepackage{graphicx}

\title{Finding active galactic nuclei through Fink}

\author{%
  Etienne Russeil\\
  Université Clermont Auvergne\\
  CNRS/IN2P3, LPC\\
  F-63000 Clermont-Ferrand, France\\
  \texttt{etienne.russeil@uca.fr}
  \And
  Emille E. O. Ishida\\
  Université Clermont Auvergne\\
  CNRS/IN2P3, LPC\\
  F-63000 Clermont-Ferrand, France\\
  \texttt{emille.ishida@clermont.in2p3.fr}
  \And
  Roman Le Montagner\\
  Université Paris-Saclay\\
  CNRS/IN2P3, IJCLab\\
  91405 Orsay, France\\
  \texttt{roman.le-montagner@ijclab.in2p3.fr}
  \And
  Julien Peloton\\
  Université Paris-Saclay\\
  CNRS/IN2P3, IJCLab\\
  91405 Orsay, France\\
  \texttt{peloton@ijclab.in2p3.fr}
  \And
  Anais Möller\\
  Centre for Astrophysics and Supercomputing\\
  Swinburne University of Technology\\
  Mail Number H29, PO Box 218, 31122, Hawthorn, VIC, Australia\\
  \texttt{amoller@swin.edu.au}
}

\begin{document}

\maketitle

\begin{abstract}
We present the Active Galactic Nuclei (AGN) classifier as currently implemented within the \fink\ broker. Features were built upon summary statistics of available photometric points, as well as color estimation enabled by symbolic regression. The learning stage includes an active learning loop, used to build an optimized training sample from labels reported in astronomical catalogs. Using this method to classify real alerts from the Zwicky Transient Facility (ZTF), we achieved 98.0\% accuracy, 93.8\% precision and 88.5\% recall. We also describe the modifications necessary to enable processing data from the upcoming Vera C. Rubin Observatory Large Survey of Space and Time (LSST), and apply them to the training sample of the Extended LSST Astronomical Time-series Classification Challenge (ELAsTiCC). Results show that our designed feature space enables high performances of traditional machine learning algorithms in this binary classification task. 
\end{abstract}

\section{Introduction}

Active Galactic Nuclei (AGN) are bright, variable astrophysical sources associated with the inflow of circumstellar matter into central galactic black holes \cite{padovani2017}. From the observer perspective, they comprise a large set of light curve behaviors, including instances where observational patterns evolve with time \cite{Noda2018, sanchez2021}. Beyond being paramount for the study of accretion and photoionization physics \cite{Trakhtenbrot2019}, they can trace star formation regions \cite{masoura2018} and have the potential to enrich cosmological studies \cite{martinez2019}. Thus, reliable and cheap identification of large populations of AGNs are crucial to enable a better understanding of their mechanisms and their impact in the galactic environment. 

The Vera C. Rubin Observatory Legacy Survey of Space and Time\footnote{\url{https://www.lsst.org/}} (LSST), expected to start operations in 2024, will produce a large volume of photometric data, including a diverse AGN population. Each time a brightness variability beyond 5-$\sigma$ from the background is detected, an alert will be generated. We expect 10 million of such alerts per night, which will be streamed to chosen community brokers. \fink
\footnote{\url{https://fink-broker.org/}}, is one of the official LSST brokers, whose task is to receive this data, extract meaningful information from it and re-distribute it to scientific communities. The broker contains a series of machine learning based modules, which enables fast processing of the data stream.
In preparation for the start of LSST, \fink\ is currently operating on data from the Zwicky Transient Facility (ZTF) (\cite{Patterson_2018}), which produces around 300 000 alerts per night.

This work presents details about the AGN classifier within the \fink\ broker. It includes a tailored feature extraction procedure followed by the construction of an optimized training sample using uncertainty sampling active learning and a random forest classifier \cite{ho1995random}.

\section{Data}

\begin{figure}
    \centering
    \includegraphics[scale=0.33]{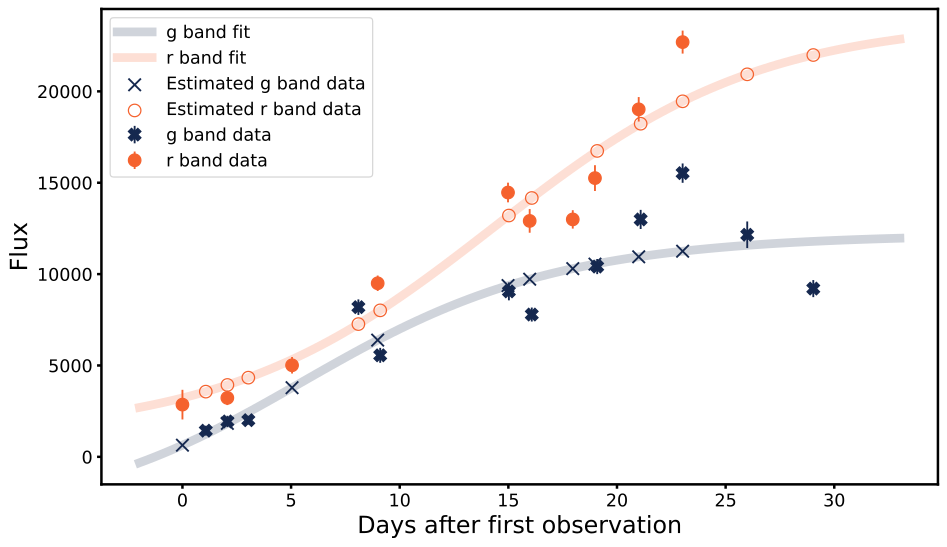}
    \caption{Example of \finkd\ light curve. The object belongs to the BLLac class, with a ZTF objectID = "ZTF18abzwaiw" and an alert ID = 1424109966315015005. Bold symbols show observed points. The continuous lines show flux estimation using Equation \ref{eq:bump} and open symbols denote flux value used for color estimation. ZTF filters $g$ and $r$ are shown in dark blue and orange, respectively. }
    \label{fig:lc}
\end{figure}

Two databases were used in this work. \finkd\ corresponds to all ZTF data collected\footnote{\url{https://zenodo.org/record/5645609}} by \fink\ from Nov/2019 to Mar/2021, for which an associated label was found in \simbad\footnote{\url{https://simbad.unistra.fr/simbad/}} or in Transient Name Server (\tns\footnote{\url{https://www.wis-tns.org/}}). For each filter, the maximum flux was normalized to 1 and the time of first observation was shifted to zero. We then randomly sampled 100k AGN and 1 million non-AGNs alerts, and only considered objects with a minimum of 4 observed points in each filter. 
The resulting database contained 607772 alerts. Following \cite{Leoni_2022},
we regrouped objects into 2 larger categories, "AGN" and "non-AGN". The AGN category encloses [AGN, LINER, Blazar, BLLac, QSO]\footnote{\simbad\ tags were used before the June 2022 taxonomy modification.}.
Among \finkd, 536621 alerts belong to the non-AGNs, while 71151 are AGNs. The resulting set was then equally divided into two samples: training and testing. Since each object can produce multiple alerts, we ensured that alerts from a given object were only present in one of the two samples. The \finkd\ represents the state of the art of what can be done with real data. 

Aiming to estimate the performance of our classifier in a LSST-like data environment, we put together a second database, hereafter, \texttt{ELAsTiCC\_data} using data from the Extended LSST Astronomical Time-series Classification Challenge\footnote{\url{https://portal.nersc.gov/cfs/lsst/DESC_TD_PUBLIC/ELASTICC/}} (\elasticc). It represents a more complex version of the Photometric LSST Astronomical Time-series Classification Challenge\footnote{\url{https://www.kaggle.com/c/PLAsTiCC-2018}} (\texttt{PLAsTiCC}), held in 2018. Focusing in anticipating an LSST-like data and software environment, \elasticc\ aims to test not only the classification power of broker systems, but also the resilience of their infrastructure and link with LSST facilities. It uses 32 different transient template models to generate simulated light curves using the Supernova Analysis (\texttt{SNANA}) \cite{kessler2009} code. These are cut into alerts (each new photometric observation above detection threshold generate an alert) and daily streamed to broker teams. The brokers are expected to process the stream and send back their probability scores. Results will compare the performance of all teams who provided scores after an initial period of 3 months. A training sample composed of full light curves, and with a different cadence from the one used to create the test sample, was made available so broker teams could prepare/train their models for the challenge. This training sample is our starting point for the construction of the \texttt{ELAsTiCC\_data} database. From it, we selected 50k AGNs and 50k non-AGNs objects with at least 4 points in 2 consecutive filters for training and another 50k AGN/50k non-AGN for testing.

\section{Methodology}

\begin{figure}
    \centering
    \includegraphics[scale=0.3]{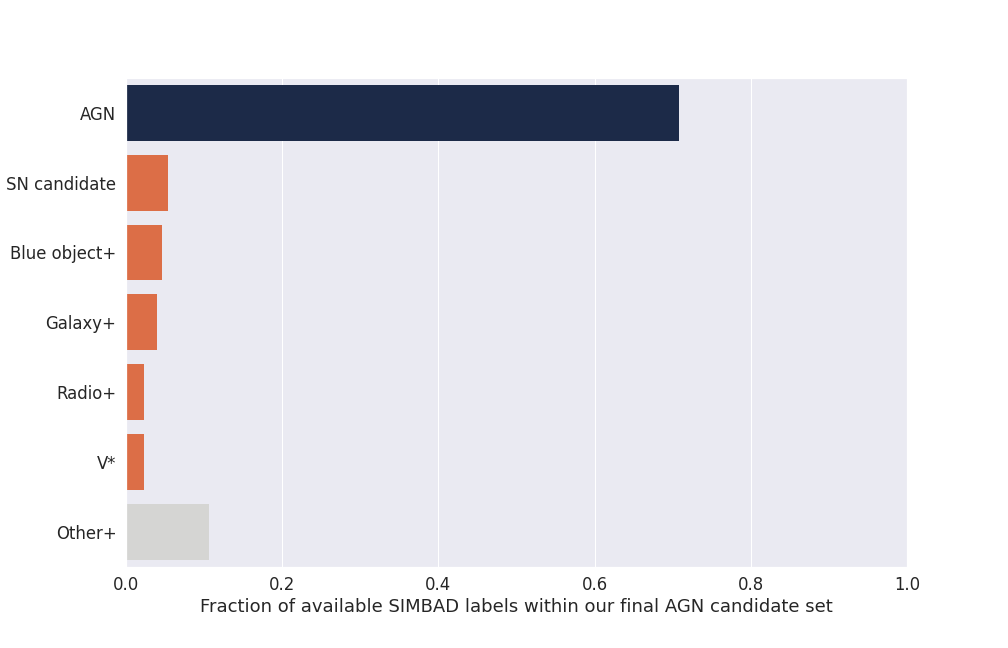}
    \caption{Broad categories of \simbad\ and \tns\ labels within the sample of identified AGN candidates. It follows \simbad\ taxonomy and the + marks classes which may include AGN sub-types.}
    \label{fig:cont}
\end{figure}

For each filter, we used the photometric points to compute the following features: maximum flux before normalization; standard deviation of the flux; number of points and mean signal to noise ratio. Moreover, we also added metadata to the features table, meaning: right ascension; declination; standard deviation and maximum absolute value of each color. Additionally, only for \texttt{ELASTiCC\_data}, we added metadata information regarding: host galaxy redshift, host galaxy redshift error and host galaxy distance to the object. 

The color calculation for this type of data is specially delicate. Since the sampling of the points is irregular, no pair of points exist at the exact same time in different filters, therefore we must interpolate each light curve so a proper colour estimation can be performed. In searching for a smooth function which would correctly capture the overall behavior of AGNs, we opted to use:

\begin{equation}
f(t) = \frac{1}{1 + e^{-At -e^{Bt} + C}} + D
\label{eq:bump}
\end{equation}

This form was obtained by applying a traditional symbolic regression algorithm to noiseless simulated transient light curves via
\texttt{gplearn}\footnote{\url{https://github.com/trevorstephens/gplearn}}. Once the code had converged and an analytical form propose, we manually substitute the float values by parameters, resulting in the reported functional form (Equation \ref{eq:bump}). Therefore parameters A, B and C do not have any direct physical interpretation. However the parameter D was added latter on and represents the flux baseline. The minimization is performed using the least squared function from the python module iMinuit (\cite{iminuit}). Figure \ref{fig:lc} shows an example of a real alert from \finkd, alongside the estimated fluxes. Color was computed by subtracting flux values between two consecutive filters, [g-r] for \finkd\ and [u-g, g-r, r-i, i-z, z-Y] for \elasticcd, at the condition that both concerned filters contains at least 4 observed points (considering forced photometry but not upper limits). 

Once the feature matrix was constructed, classification proceeded using the random forest algorithm from the python module scikit-learn (\cite{scikit-learn}). This method uses an ensemble of decision trees, each built from a different sub-sample of the data. When new data is inputted, the answer from each tree is counted as a vote, and the probability outputted by the forest comes from the proportion of each vote. In what follows, all results were obtained using 100 trees. Conducted attempts using a gradient boosted tree algorithm, as well as attempts to increase the number of trees, lead to no further improvements in the classification score.

For \finkd, an optimized training sample was built by using an active learning strategy based on uncertainty sampling \cite{ishida2019}. We started the learning cycle with 5 AGN and 5 non-AGNs randomly selected from the sample of available labels, and trained a random forest using these 10 objects while the rest of the data was used for testing. The alert in the testing sample for which the random forest is the most uncertain about (probability closest to 50\% percent) is moved from testing sample to training sample in each cycle. We performed 2000 loops, resulting in a final training sample of 2010 objects. The entire process was performed 10 times in order to briefly access the impact of different initial conditions. In the case of \elasticcd, representativeness was ensured by construction, thus no active learning loop was required.

\section{Results and conclusion}

\begin{figure}
    \centering
    \includegraphics[scale=0.3]{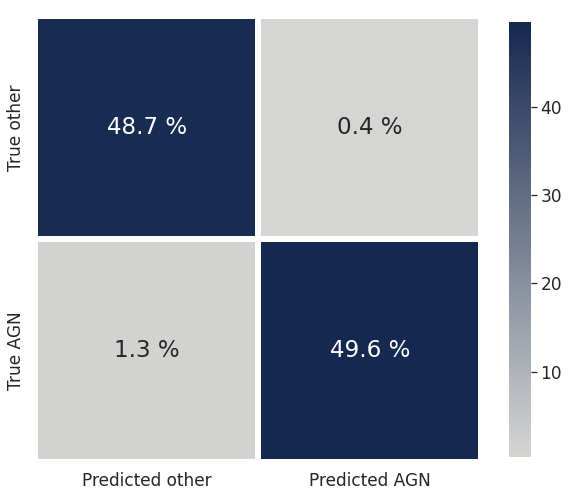}
    \caption{Confusion matrix results from the \elasticcd\ sample.}
    \label{fig:conf}
\end{figure}

The active learning process will, by construction, build a training data set approximately made of 50\% AGNs and 50\% non-AGNs. In the case of unbalanced data set such as \finkd, this property is crucial. A balanced and informative training set ensures that the random forest is learning to separate types of alert, and not just learning the statistical distribution of this particular dataset. Furthermore a balanced training dataset guarantees the classifier optimal threshold to be around 0.5.

When applied to all the non queried objects, the classifiers built with active learning achieved on average 98.6 \% accuracy, 95.9 \% precision and 91.9 \% recall on finding AGNs, with a standard deviation of 0.1\%, 0.7\% and 0.8\% respectively. 

We conclude that the uncertainty sampling was efficient at building a training sample. It also ensures that the classifier will perform well when used on another dataset with different statistical properties. Among the 10 runs, we kept the classifier with highest product recall $\times$ score, which in our case also satisfies the F1 criteria. Applying this classifier to the testing sample (not involved in training), we achieve a score of 98.0 \% accuracy with 93.8 \% precision and 88.5 \% recall on AGNs. Finally the classifier has been tested directly on the Fink alert stream. All alerts between Jan/2019 to Jan/2020 were processed and we evaluate the results for every labelled alerts (excluding objects used in the training sample). Figure \ref{fig:cont} shows the distribution of labels within the photometrically predicted AGNs. Results from the \elasticcd\ sample are shown in Figure \ref{fig:conf}. Feature importance analysis on results obtained from the \finkd\ shows that all features play a similar role in enabling classification. For the \elasticcd, the number of points acts like a first layer of classification. This was expected, since this data set contain a high percentage of transient classes. These are objects whose brightness are only significant during a limited time window, which makes them more easily distinguishable from AGNs, that are persistent sources.

Both models described in this work are now integrated to the \fink\ broker, lively processing alerts from ZTF and \elasticc. The filtered ZTF stream will be directed to the AGN community interested in spectroscopic follow-up of individual events. We foresee the development of similar models focused on sub-types of AGNs chosen by \fink\ users. Finally, we intend to use the AGN scores to further filter out the data given to the supernova classifiers in \fink, thus allowing for an even higher precision in the estimates delivered by the broker.

All the work presented in this paper is publicly available on GitHub \footnote{\label{foot:git} \url{https://github.com/astrolabsoftware/fink-science/tree/master/fink_science/agn}}.

\section*{Acknowledgements}
We thank David O. Jones for making simulations available for this project. This work was developed within the \fink\ community and made use of the \fink\ community broker resources. \fink\ is supported by LSST-France and CNRS/IN2P3.

\bibliography{main}

\end{document}